\documentclass[aps,prb,twocolumn,groupedaddress]{revtex4}
\usepackage{graphics}
\usepackage{graphicx}

\begin{document}


\title{Stretched exponential relaxation and ac universality in disordered dielectrics}


\author{Alexander V. Milovanov}
\email[]{Alexander.Milovanov@phys.uit.no}

\altaffiliation{On leave from: Department of Space Plasma Physics, Space Research Institute, Russian Academy of Sciences, Profsoyuznaya 84/32, 117997 Moscow, Russia}
\affiliation{Department of Physics and Technology, University of Troms\o, N-9037 Troms\o, Norway}

\author{Jens Juul Rasmussen}
\email[]{jens.juul.rasmussen@risoe.dk}
\affiliation{Optics and Plasma Research Department, Ris\o\,National Laboratory, Technical University of Denmark, Building 128, P.O. Box 49, DK-4000 Roskilde, Denmark}

\author{Kristoffer Rypdal}
\email[]{Kristoffer.Rypdal@phys.uit.no}

\affiliation{Department of Physics and Technology, University of Troms\o, N-9037 Troms\o, Norway}



\begin{abstract}
This paper is concerned with the connection between the properties of dielectric relaxation and ac (alternating-current) conduction in disordered dielectrics. The discussion is divided between the classical linear-response theory and a self-consistent dynamical modeling. The key issues are, stretched exponential character of dielectric relaxation, power-law power spectral density, and anomalous dependence of ac conduction coefficient on frequency. 
We propose a self-consistent model of dielectric relaxation, in which the relaxations are described by a stretched exponential decay function. Mathematically, our study refers to the expanding area of fractional calculus and we propose a systematic derivation of the fractional relaxation and fractional diffusion equations from the property of ac universality.  
\end{abstract}


\maketitle

\section{Introduction}

Many materials with a disordered structure show a dielectric relaxation that is not described by an exponential (i.e., Debye-like) decay with a characteristic single decay time. Rather the relaxations follow a stretched exponential the so-called Kohlrausch-Williams-Watts (KWW) function \cite{Kohlrausch,Williams}
\begin{equation}
\phi _{\beta} (t) = \exp [-(t/\tau)^\beta]
\end{equation}
with the exponent $0 < \beta \leq 1$ and $\tau$ a constant. The $\beta$ values generally depend on the absolute temperature and the chemical composition of the material and typically span between 0.3 and 0.8. The relaxation pattern in Eq. (1) has been found empirically in various amorphous materials as for instance in many polymers and glass-like materials near the glass transition temperature (for review see Refs. \onlinecite{Kaatz} and \onlinecite{Phillips}, and references therein). Some physical models yielding general features of the KWW type dielectric relaxation are discussed in Refs. \onlinecite{Phillips} and \onlinecite{Montroll} where one also finds a review of experimental dielectric relaxation data. Other applications of the KWW relaxation function include long-time decay in trapping processes, \cite{Grassberger} non-radiative exciton recombination, \cite{Halas} and  relaxation in sand-piles. \cite{Tang} 

The stretched exponential KWW decay function $\phi _{\beta} (t)$ can conveniently be considered as a weighted average of single-exponential relaxation functions 
\begin{equation}
\phi _{\beta} (t) = \int _0 ^\infty \varrho _\beta (\mu) \exp (-t/\mu) d\mu
\end{equation}
where the weighting function $\varrho _\beta (\mu)$ is expressible in terms of a stable (L\'evy) distribution \cite{Montroll} (Eq.~(\ref{Levy}) below). Because of this connection with the statistics of stable laws, the KWW relaxation properties can be argued to appear naturally through the dynamics, thus challenging the common view that the stretched exponential function is just a suitable phenomenological fitting tool without fundamental significance. \cite{School} We draw attention to the fact that the KWW relaxations arise from a superposition of many single-exponential relaxation processes and are multi-scale, contrary to a Debye relaxation.   

In this paper we are concerned with the connection between the KWW relaxation function in Eq. (1) and the anomalous frequency dependence of ac conductivity of homogeneously disordered insulators. It has been found also partly empirically that insulating and/or poorly conducting materials with molecular and/or structural disorder exhibit a common conductivity (dielectric) response, a phenomenon often referred to as ``ac universality" (see review \onlinecite{Dyre}, and original works in Refs. \onlinecite{Rolla} and \onlinecite{Jacobs}). This response is characterized by low-frequency conductivity with very weak or no frequency dependence, and a higher-frequency counterpart that follows an approximate power law 
\begin{equation}
\sigma^{\prime} (\omega) \propto \omega ^\eta \label{1.3}
\end{equation}
with $\sigma^{\prime} (\omega)$ the real part of the frequency-dependent complex conductivity $\sigma (\omega)$ and the exponent $\eta$ ranging between 0 and 1, and most often between 0.6 and 1, depending on the material and the absolute temperature. The defining feature of ac universality is independence of the microscopic details of the disorder and of the nature of the charge conduction mechanism operating in the system (classical barrier crossing for ions and/or quantum mechanical tunneling for electrons). Note that the power-law in Eq.~(\ref{1.3}) is well-defined for frequencies higher than the dielectric loss peak frequency. The system-specific properties are contained in the coefficient in front of $\sigma^{\prime} (\omega)$ (not shown in the scaling relation). The signatures of ac universality have been found in materials as diverse as ion conducting glasses, amorphous and polycrystalline semiconductors, organic-inorganic composites, ion and electron conducting polymers, and doped semiconductors at helium temperatures  (Ref. \onlinecite{Dyre} for a review). The observed characteristics including the temperature dependence of the exponent $\eta$ could be reproduced in a model \cite{PRB01} in which the conductivity is caused by random motion of charged-particles on a fractal lattice. \cite{Gefen} The scale-free behavior of the ac conductivity at higher frequencies could be explained as arising from the hierarchic structure of the fractal (see Fig. 1) in which multiple cycles and dead ends produce effective potential traps influencing the motion of  the charge carriers. This fractal-based approach conforms in spirit with the issue of trap-controlled conduction and multiple-trapping transport of charges in disordered media (Ref. \onlinecite{Anta} and references therein).
\begin{figure}
\caption{\label{} Schematic of cycles and dead-ends of a percolating fractal structure.}
\end{figure}

The purpose of this paper is to explore the connection between ac universality and the stretched exponential KWW type dielectric relaxation in disordered dielectrics. We shall argue that the two phenomena share the statistical-mechanical foundation, and a simple relation between the exponents $\beta$ and $\eta$ will be derived:
\begin{equation}
\beta + \eta = 1
\end{equation}
The paper is organized as follows. In Sec. II, we show that a KWW-type dielectric relaxation implies a power-law memory response function for short times and, related to this feature, high-frequency dependence of ac conduction coefficient with the exponent $\eta = 1 - \beta$. We then discuss the route to the power spectrum and the occurrence of power-laws for high frequencies in the power spectral density (PSD). As a consistency check, we re-derive the power spectrum from a Pareto-L\'evy distribution of relaxation times and demonstrate that the power-law spectrum is directly related to a power-law distribution of relaxation times for the short time scales. In Sec. III, we present a self-consistent model of dielectric relaxation, in which the polarization and electric source fields are self-consistently generated by the residual polarization-charge density. We show that, if the dielectric observes ac universality, the relaxations are stretched-exponential for short times. Mathematically, our study refers to the expanding area of fractional calculus and we propose a systematic derivation of the fractional relaxation and fractional diffusion equations from the property of ac universality. We summarize our conclusions in Sec. IV.  

\section{Response functions, scaling, and power spectral density}

\subsection{Response functions}\label{sec:response}
Let a homogeneous, isotropic dielectric be exposed to the external polarizing electric field ${\bf E} = {\bf E} (t, {\bf {r}})$, which we consider as a function of time $t$ and the space coordinate, ${\bf r}$. By homogeneous and isotropic we refer to spatial scales larger than the typical scales of the molecular/structural disorder. Assuming a linear and spatially local response of the material the polarization field at time $t$ at point ${\bf {r}}$ can be written as    
\begin{equation}
{\bf P} (t, {\bf {r}}) = \int _{-\infty}^{+\infty} \chi (t - t^{\prime}) {\bf E} (t^{\prime}, {\bf {r}}) dt^{\prime}\label{2.1} 
\end{equation}
where $\chi (t - t^{\prime})$ is a response or memory function. Causality requires that $\chi(t - t^{\prime}) = 0$ for $t < t^{\prime}$. By considering the source field of the form ${\bf E}(t,{\bf r})={\bf E}({\bf r})\delta (t)$ with $\delta (t)$ the Dirac delta function we have the polarization response ${\bf P}(t,{\bf r})={\bf E}({\bf r})\chi(t)$, hence $\chi(t)$ is the response to a delta-pulse in the source field. 

In a basic theory of the dielectric relaxation one is interested in the polarization response to a field which is steady for $t < 0$ and, then, is suddenly removed at time $t=0$. One introduces the relaxation function $\phi(t)$ as the magnitude of the polarization response on an electric field with time history ${\bf E}(t,{\bf r}) = {\bf E} ({\bf r})\theta (-t)\exp{(\nu t)}$ in the limit $\nu\rightarrow +0$, where $\theta(t)$ is the Heaviside step function. The infinitely slow exponential growth $\exp{(\nu t)}$ is included in order to satisfy ${\bf E}(t,{\bf r}) = 0$ for $t\rightarrow -\infty$, required by causality. From Eq.~(\ref{2.1}) one can see that the polarization response is ${\bf P}(t,{\bf r})={\bf E}({\bf r})\phi (t)$, where 
\begin{equation}
\phi (t) = \theta(-t)\phi(0)\exp{(\nu t)} + \theta(t)\left(\phi(0)-\int_0^t\chi (t^{\prime}) dt^{\prime}\right)\label{2.2}
\end{equation}
and we defined
\begin{equation}
\phi(0) = \int_0^{\infty} \chi(t^{\prime}) dt^{\prime}\label{2.2a}
\end{equation}
In the literature one commonly writes the value of $\phi(0)$ in terms of the static permittivity of the dielectric medium $\epsilon (0)$, i.e., $\phi(0) = \left[\epsilon (0) - 1\right] /4\pi$. The definition in Eqs.~(\ref{2.2}) and~(\ref{2.2a}) is practical for exploring the connection between the memory kernel and the relaxation function. For these practical reasons, we shall set $\phi(0) = 1$ in the calculation below. Performing $d / dt$ on Eq.~(\ref{2.2}) one obtains
\begin{equation}
\chi (t) = -\frac{d\phi}{dt} \label{2.3}
\end{equation}
If we postulate the relaxation function in the KWW form 
\begin{equation}
\phi (t) = \theta(-t)\exp{(\nu t)} + \theta(t)\exp{[-(t/\tau)^{\beta}]} \label{KWW}
\end{equation}
from Eq.~(\ref{2.3}) we find
\begin{equation}
\chi (t) = \frac{\beta}{\tau^{\beta}}t^{\beta-1} \theta(t) \exp{[-(t/\tau)^{\beta}]} \label{2.4}
\end{equation}
This is a very important expression, because it reveals that the stretched exponential relaxation function $\phi(t)$ leads to a response function $\chi(t)$ which behaves like a power law $t^{\beta-1}$ on short time scales $t\lesssim\tau$, with a stretched exponential cut-off on long time-scales $t\gtrsim\tau$. This cut-off is necessary for the existence of  a finite relaxation function as a response to the step-function electric field ${\bf E}(t,{\bf r}) = {\bf E} ({\bf r})\theta (-t)\exp{(\nu t)}$, since without it the integrals in Eqs.~(\ref{2.1}) and~(\ref{2.2a}) will diverge. However, if the electric field is an oscillating function or a white noise, the response given by Eq.~(\ref{2.1}) will converge even with $\chi(t)\sim t^{\beta-1}$  without such a cut-off. The response corresponding to such an ``unscreened'' power-law response function belongs to the class of  fractional Brownian functions, \cite{Isi} and will be discussed in some detail in Sec.~\ref{sec:PSD}.
By Fourier transforming Eq.~(\ref{KWW}) we have 
\begin{equation}
\phi (\omega) = \int_{-\infty}^0 \exp(\nu t) e^{i\omega t} dt + \int_0^{+\infty} \exp [-(t/\tau)^{\beta}] e^{i\omega t} dt \label{FT}
\end{equation}
yielding, for $\nu\rightarrow +0$ and $\omega$ higher than a non-vanishing lower bound (physically corresponding to the dielectric loss peak frequency):
\begin{equation}
\phi (\omega) = \tau\mathrm{Q} (\omega \tau) + i\tau\left(\mathrm{V} (\omega \tau) - \frac{1}{\omega\tau}\right)\label{FFT}
\end{equation}
where $\mathrm{Q}$ and $\mathrm{V}$ are the L\`evy definite integrals: 
\begin{equation}
\mathrm{Q} (z) = \int_0^{+\infty}\exp{(-u^{\beta})} \cos{(uz)}du \label{QQ}
\end{equation}
\begin{equation}
\mathrm{V} (z) = \int_0^{+\infty}\exp{(-u^{\beta})} \sin{(uz)}du \label{VV}
\end{equation}
here expressed as functions of dimensionless frequency $z = \omega\tau$.

\subsection{Leading-term approximation}
Series expansion of the L\'evy functions $\mathrm{Q} (z)$ and $\mathrm{V} (z)$ has been derived and discussed in the literature. \cite{Kaatz,Montroll,Wintner} Here we utilize an expansion good for higher frequencies, which goes in inverse powers of $z$:
\begin{equation}
\mathrm{Q} (z) = \sum _{n=1}^\infty (-1)^{n-1} \frac{1}{z^{n\beta + 1}} \frac{\Gamma (n\beta + 1)}{\Gamma (n + 1)} \sin \frac{n\beta\pi}{2}\label{SQQ}
\end{equation} 
\begin{equation}
\mathrm{V} (z) = \sum _{n=0}^\infty (-1)^{n} \frac{1}{z^{n\beta + 1}} \frac{\Gamma (n\beta + 1)}{\Gamma (n + 1)} \cos \frac{n\beta\pi}{2}\label{SVV}
\end{equation}    
From Eqs.~(\ref{SQQ}) and~(\ref{SVV}) one can see that the expansion of $\mathrm{Q} (z)$ starts from a term which is proportional to $z ^{-(1+\beta)}$ and so does the expansion of $\mathrm{V} (z) - 1/z$. Hence, up to higher order terms, $\phi (\omega) \propto \omega ^{-(1+\beta)}$. 
 
If we now define the frequency-dependent complex susceptibility of the dielectric, $\chi(\omega)$ as a Fourier pair with $\chi(t)$, from Eqs.~(\ref{2.3}) and~(\ref{FFT}) we find 
\begin{equation}
\chi (\omega) = i\omega \phi (\omega) = 1 - \omega \tau \mathrm{V} (\omega \tau) + i\omega \tau\mathrm{Q} (\omega \tau)\label{Chi}
\end{equation}
with the leading term $\chi(\omega) \propto \omega ^{-\beta}$. With use of the Kramers-Kronig relation $\chi(\omega) \propto \mathrm{P} \int d\omega^{\prime} \sigma (\omega^{\prime}) / \omega^{\prime} (\omega^{\prime} - \omega)$ the scaling of the frequency-dependent complex conductivity of the material can be evaluated as  
\begin{equation}
\sigma(\omega) \propto \omega^{1-\beta} \label{Sigma}
\end{equation}
which reproduces the phenomenological form in Eq. (3) with $\eta = 1-\beta$, whence Eq. (5) is obtained. 

\subsection{Power spectral density}\label{sec:PSD}
By Fourier transforming Eq.~(\ref{2.1}) we find the following expression for the PSD of the polarization field:
\begin{equation}
\mathrm{S}(\omega) = \langle |{\bf P}(\omega, {\bf {r}})|^2\rangle=|\chi(\omega)|^2\langle |{\bf E}(\omega, {\bf {r}})|^2\rangle\label{2.10}
\end{equation}
where the angle brackets $\langle \cdot \rangle$ denote an ensemble average. From Eq.~(\ref{2.10}) one can see that $|\chi(\omega)|^2$ is the PSD of the polarization field when the driving electric field is an uncorrelated white noise signal. In Fig.~\ref{chispectra} we have plotted (in log-log axes) the $|\chi(\omega)|^2$ function from Eq.~(\ref{Chi}), which shows a transition to a power-law regime for $\omega\tau \gtrsim 1$.  
\begin{figure}
\caption{\label{chispectra} The square amplitude of the frequency-dependent complex susceptibility vs the normalized frequency, $\omega\tau$, for different values of the relaxation exponent $\beta$. The plots show a transition to a power-law regime for $\omega\tau \gtrsim 1$.}
\end{figure}

The origin of the power-law regime at the high frequencies can be found in the short-time behavior of $\chi(t)$ which diverges as $t^{\beta - 1}$ when $t\rightarrow 0$. In fact, the Fourier integral 
\begin{equation}
\chi(\omega) = \frac{\beta}{\tau^\beta} \int_0^{\infty} t^{\beta - 1}\exp{\left[-(t/\tau)^{\beta} + i\omega t\right]} dt \label{LL}
\end{equation}
converges for $t/\tau \rightarrow 0$ if $\beta > 0$. This integral is essentially the Fourier transform of the power-law function $t^{\beta - 1}$ with a stretched exponential cut-off starting to take effect in the integrand for time scales $t\sim \tau$ or longer (i.e., the frequencies $\omega \sim 2\pi / \tau$ or smaller). For $\omega \tau$ sufficiently large the integral has already nearly converged before the cut-off region in the integrand is reached, enabling one to evaluate the susceptibility function as  
\begin{equation}
\chi(\omega) \approx \frac{\beta}{\tau^\beta} \int_0^{\infty} t^{\beta - 1} \exp{(i\omega t)} \,dt \propto \omega^{-\beta}\label{LLL}
\end{equation}
The scaling in Eq.~(\ref{LLL}) complies with the leading term in the series expansion of the complex function $1 - z\mathrm{V} (z) + iz\mathrm{Q} (z)$. 

From Eq.~(\ref{2.10}) one obtains, for the correspondingly high frequencies, $\mathrm{S}(\omega) \propto \omega ^{-2\beta}$ provided that the input driving field is an uncorrelated white noise signal. 

The short-time power-law behavior $\chi (t)\sim t^{\beta-1}$ corresponds to a memory kernel which defines, by means of a convolution with Gaussian white noise, the family of fractional Brownian functions, \cite{Isi} i.e., $\chi_H (t) \sim t^{H-1/2}$ where the exponent $H$ is often referred to as the Hurst exponent. The connection with the KWW relaxation function exponent is given by
\begin{equation}
H = \beta - 1/2 \label{2.14}
\end{equation}
Depending on the value of $H$, one distinguishes between fractional Brownian motions ($0 < H < 1$) and fractional Brownian noises ($H < 0$). The difference is that a noise-like function has a stationary quality in the sense its variance does not asymptotically grow with time. From Eq.~(\ref{2.14}) one can see that on time scales shorter than $\tau$ the response signal on a white noise source field is a persistent fractional Brownian noise for $0<\beta<1/2$ (since $-1/2<H<0$) and an antipersistent fractional Brownian motion  for $1/2<\beta<1$ (since $0<H<1/2$). In the Debye limit of $\beta\rightarrow 1$, the response signal is an ordinary Brownian motion with $H = 1/2$. Note that $\beta = 1/2$ corresponding to $H = 0$ is a special value which separates motion-like and noise-like response processes. 

On time scales longer than $\tau$, the memory function $\chi (t)$ is strongly influenced by the stretched exponential decaying factor. For $\beta < 1$ the long stretched exponential range leads to a deviation from the flat white noise spectrum for the low frequencies $\omega \ll 2\pi / \tau$ and more so for smaller $\beta$. This implies correlations on time scales $t\gg \tau$ (though not yet long-range correlations \cite{Beran} in the strict sense, since the latter would require a spectrum which diverges faster than logarithmic as $\omega \rightarrow 0$).

From Eq.~(\ref{2.3}) we have $\phi(\omega) = -i\chi(\omega)/\omega$. Hence, the power spectrum of the response to a step function driving field is
\begin{equation}
\mathrm{S} (\omega) = |\phi(\omega)|^2 = \omega^{-2} |\chi(\omega)|^2  \propto \omega^{-2(1+\beta)} \label{2.16}
\end{equation}
The extra factor $\omega^{-2}$ in the PSD compared to the PSD with an uncorrelated white noise driving field is due to the non-stationarity of the step function source signal. 

\subsection{The physical origin of power-law spectra and stretched exponential relaxation}
It is instructive to re-derive the power spectrum in Eq.~(\ref{2.16}) from the distribution of relaxation times as characterized by Eq. (2), then from the weighted superposition of the Debye single-exponential relaxation processes. From Eq.~(\ref{FT}) we have 
\begin{equation}
\phi (\omega) = -\frac{i}{\omega} + \int _{0}^{\infty} e^{i\omega t} dt \int _0 ^\infty \varrho _\beta (\mu) \exp (-t/\mu) d\mu\label{Exp}
\end{equation}
where we expanded the stretched exponential over the partial relaxation times, $\mu$. Changing order of integration in Eq.~(\ref{Exp}) and Fourier transforming the exponential function $\exp (-t/\mu)$ one finds
\begin{equation}
\phi (\omega) = -\frac{i}{\omega} + \int _0 ^\infty \frac{\mu}{1-i\omega\mu} \varrho _\beta (\mu) d\mu\label{Fexp} 
\end{equation}
The weighting function $\varrho _\beta (\mu)$ is given by Eqs. (51d) and (55) of Ref. \onlinecite{Montroll} where one replaces the exponent $\alpha$ with $\beta$, the time constant $T$ with $\tau$, and the variable $\mu$ with $\tau / \mu$. In our notations:
\begin{equation}
\varrho _\beta (\mu) = \frac{\tau}{\mu ^2}\, \mathrm{L} _{\beta, -1} (\tau / \mu)\label{Levy}
\end{equation} 
where $\mathrm{L} _{\beta, -1}$ is the L\'evy distribution function with  skewness parameter $-1$ (Ref. \onlinecite{Wolf}).  For short relaxation times $\mu \ll \tau$ the parameter $\tau/\mu$ corresponds to the tail of the L\'{e}vy  distribution, which is approximated by  the Pareto inverse-power distribution, \cite{Wolf} i.e.,  
\begin{equation}
\mathrm{L} _{\beta, -1} (\tau / \mu) \approx A _\beta (\tau / \mu) ^{-(1+\beta)}\label{Pareto}
\end{equation} 
where $A _\beta$ is a normalization parameter and  $0<\beta<1$. The distribution of relaxation times is then a pure power law:
\begin{equation}
\varrho _\beta (\mu) \propto \mu^{-2}\mu^{1+\beta} \propto \mu ^{-(1-\beta)} \label{Pareto1}
\end{equation}
Now let us demonstrate directly the connection between the power-law distributed relaxation times for $\mu \ll \tau$ and the power-law power spectra for $\omega \gg 2\pi / \tau$. By considering $\omega \gg 2\pi / \tau$ we can define a cross-over time scale $a \sim \Theta / \omega$ with $\Theta$ a constant much larger than 1.  We can then split the integration in Eq.~(\ref{Fexp}) into a tail integral  $\int_0^{a} d\mu$ and a core integral $\int_a^{\infty} d\mu$. In the core integral we have $\omega \mu \geq \Theta \gg 1$ and  we can neglect 1 in $1-i\omega\mu$, then integrate the distribution function $\varrho _\beta (\mu)$ through all $\mu$ to find, with $\int \varrho _\beta (\mu) d\mu = 1$, 
\begin{equation}
\int _a^{\infty} \frac{\mu}{1-i\omega\mu} \varrho _\beta (\mu) d\mu \approx \frac{i}{\omega}\label{Comp}
\end{equation}
Thus the core integral compensates for $-i/\omega$ in Eq.~(\ref{Fexp}), leading to
\begin{equation}
\phi (\omega) \approx \int _0^a \frac{\mu}{1-i\omega\mu} \varrho _\beta (\mu) d\mu\label{TT} 
\end{equation}
In  the tail integral we have $\tau/\mu \gg 1$, and we can  substitute $\varrho _\beta (\mu)$ from Eq.~(\ref{Levy}) and utilize the power-law distribution in Eq.~(\ref{Pareto1}) to obtain
\begin{equation}
\phi (\omega) \propto \int _0^a\frac{\mu ^\beta}{1-i\omega\mu} d\mu \approx \omega ^{-(1+\beta)} \int _0^{\Theta}\frac{\xi ^\beta}{1-i \xi} d\xi  \label{2.19} 
\end{equation}
with $\xi = \omega \mu$ varying from 0 to $\Theta$. The integral on the right of Eq.~(\ref{2.19}) converges to a constant, while the scaling factor in front yields the PSD in Eq.~(\ref{2.16}).

Our findings so far can be summarized as follows: The stretched exponential relaxation corresponds to a power-law memory response function for short time scales $t\lesssim\tau$. For a white noise driving electric field this in turn implies a fractional  Brownian polarization response signal on these time scales. These features are intimately connected with the description of the stretched exponential relaxation as a superposition of exponentially decaying signals with a distribution of characteristic relaxation times $\mu$ which are power-law distributed for $\mu \ll \tau$.

Power-law distribution of durations of relaxation events as responses to external perturbation is a hallmark of systems in states of self-organized criticality. \cite{Jensen} The discussion presented in this section supports the hypothesis that dielectrics exhibiting stretched exponential relaxations are in a self-organized critical state.

\subsection{Fractional-derivative representation}
If one wishes to calculate the spectrum $\mathrm{S} (\omega)$ from the dynamics of the relaxation process, the procedure is to replace the memory function in Eq.~(\ref{2.1}) by its KWW representative in Eq.~(\ref{2.4}) to give         
\begin{equation}
{\bf P} (t, {\bf {r}}) = \frac{\beta}{\tau^\beta} \int _{-\infty}^{t} dt^{\prime}\frac{{\bf E} (t^{\prime}, {\bf {r}})}{(t - t^{\prime})^{1 -\beta}} e ^{- \left[(t - t^{\prime}) / \tau \right] ^\beta}   \label{2.E1}
\end{equation}
The integration in Eq.~(\ref{2.E1}) is singular  at $t^{\prime} = t$, but the integral converges if $0<\beta<1$. As mentioned in Sec.~\ref{sec:response} the stretched exponential cut-off factor in the integrand is necessary for convergence of the integral if ${\bf E} (t^{\prime}, {\bf {r}})$ is a step function. However, for a white noise electric field the integral converges even if this cut-off is removed by replacing the exponential factor with unity. The  effect this removal will have on the nature of the response signal is to introduce long-range statistical dependence on time scales longer than $\tau$, while the  statistical properties  on scales shorter than $\tau$ are unaffected. Focusing on the shorter time scales we shall proceed with the cut-off factor removed. By applying $\partial / \partial t$ to both sides of Eq.~(\ref{2.E1}) we have
\begin{equation}
\Upsilon _\beta \frac{\partial}{\partial t}{\bf P} (t, {\bf {r}}) = {_{-\infty} \mathrm{D}}_t^{1 -\beta} {\bf E} (t, {\bf {r}})   \label{2.E2}
\end{equation}
with $\Upsilon _\beta = \tau ^\beta / \Gamma (1+\beta)$ a constant and ${_{-\infty} \mathrm{D}}_t^{1 -\beta}$ the so-called Riesz fractional derivative, \cite{Klafter} which is defined through
\begin{equation}
{_{-\infty} \mathrm{D}}_t^{1 -\beta} f (t, {\bf {r}}) = \frac{1}{\Gamma (\beta)}\frac{\partial}{\partial t} \int _{-\infty}^{t} dt^{\prime}\frac {f (t^{\prime}, {\bf {r}})}{(t - t^{\prime})^{1 -\beta}} \label{2.E3}
\end{equation}
where $f (t, {\bf {r}})$ is differintegrable for $t^{\prime} \rightarrow t$ at point ${\bf {r}}$. Along with the Riemann-Liouville derivative (to be discussed in some detail below) the Riesz derivative in Eq.~(\ref{2.E3}) offers a fractional generalization of ordinary derivative $\partial / \partial t$, which can be thought as a special case of ${_{-\infty} \mathrm{D}}_t^{1 -\beta}$ with the integer $\beta = 0$ and $1 - \beta = 1$. A Fourier transformed Riesz derivative ${_{-\infty} \mathrm{D}}_t^{1 -\beta}$ is $(-i\omega) ^{1-\beta}$, in analogy to the transform of $\partial / \partial t$. Performing a Fourier transform of Eq.~(\ref{2.E2}) we get
\begin{equation}
-i\omega {\bf P} (\omega, {\bf {r}}) \propto (-i\omega)^{1-\beta} {\bf E} (\omega, {\bf {r}}) \label{2.E4}
\end{equation}   
where we omitted $\Upsilon _\beta$ for simplicity. From Eq.~(\ref{2.E4}) one immediately recovers the PSD of the polarization field. In the integrated form Eq.~(\ref{2.E2}) reads   
\begin{equation}
\Upsilon _\beta {\bf P} (t, {\bf {r}}) = {_{-\infty} \mathrm{D}}_t^{-\beta} {\bf E} (t, {\bf {r}}) \label{2.E5}
\end{equation}
with ${_{-\infty} \mathrm{D}}_t^{-\beta}$ acting as 
\begin{equation}
{_{-\infty} \mathrm{D}}_t^{-\beta} f (t, {\bf {r}}) = \frac{1}{\Gamma (\beta)} \int _{-\infty}^{t} dt^{\prime}\frac{f (t^{\prime}, {\bf {r}})}{(t - t^{\prime})^{1 -\beta}} \label{2.E6}
\end{equation}
The operator in Eq. (\ref{2.E6}) is known as the Riesz fractional integral. With $\beta = H + 1/2$ and ${\bf E} (t, {\bf {r}})$ a Gaussian white noise the integration in Eq. (\ref{2.E5}) generates a fractional Brownian function (i.e., antipersistent fractional Brownian motion for $1/2 < \beta < 1$ and persistent fractional Brownian noise for $0 < \beta < 1/2$). The origin of fractional Brownian type polarization response to a white noise electric source field can be imagined as arising from non-random motion of dielectric molecules driven by the uncorrelated external forcing.  

\section{Fractional kinetic equations}

\subsection{Self-consistent dynamic-relaxation model}
We have assumed in Sec. II that the electric field ${\bf E} (t^{\prime}, {\bf {r}})$ has the external origin and we have investigated the properties of the polarization response field for different forms of the electric source field (i.e., the white noise and step function driving fields). In this section the assumption of the external origin will be relaxed and we shall consider ${\bf E} (t^{\prime}, {\bf {r}})$ as the inherent field of the polarization charges. 

We propose that if, in a dielectric medium, the property of ac universality is verified, then the decay of polarization goes as a KWW stretched exponential relaxation function. We formulate the problem as an initial-value problem for time non-local polarization field of the form
\begin{equation}
{\bf P} (t, {\bf {r}}) = {\bf P} (0, {\bf {r}}) + \int _{0}^{+\infty} \chi (t - t^{\prime}) {\bf E} (t^{\prime}, {\bf {r}}) dt^{\prime}\label{3.1} 
\end{equation}
where ${\bf P} (0, {\bf {r}})$ is the initial polarization. We shall assume that the response processes to all the external driving fields have been accomplished by time $t=0$ and we are interested in the self-consistent dynamics of relaxation when the residual polarization field is essentially the response to the residual electric field due to the polarization charges.   

Let $\rho (t, {\bf {r}})$ be the density of the polarization charges at time $t$ at point ${\bf r}$. The function $\rho (t, {\bf {r}})$ is defined as the mean density of the electric charges in a physically small volume around ${\bf r}$ such that the highly fluctuating molecular densities are averaged out. One assumes that there is a length scale separation between the microscopic molecular scales and the length scales on which the mean density varies. On length scales comparable to or shorter than the molecular scales, the dynamics of relaxation must be described statistically. We shall return to this issue in Sec. III C.   

Assume for simplicity that there are no external charges, i.e., the total charge of the medium is zero. The polarization charges are then the only source for the polarization and electric fields, i.e.,  
\begin{equation}
\nabla\cdot {\bf E} (t, {\bf {r}}) = 4\pi\rho (t, {\bf {r}})
\label{3.2} 
\end{equation}
and
\begin{equation}
\nabla\cdot {\bf P} (t, {\bf {r}}) = -\rho (t, {\bf {r}})
\label{3.3} 
\end{equation}
Hence
\begin{equation}
\nabla\cdot {\bf D} (t, {\bf {r}}) = 0
\label{3.4} 
\end{equation}
where ${\bf D} = {\bf E} + 4\pi {\bf P}$ is the electric displacement in the medium. The density of the polarization/relaxation currents is defined as the time derivative of the polarization field, i.e., ${\bf j} = \partial {\bf P} / \partial t$. From Eq.~(\ref{3.1}) we have
\begin{equation}
{\bf j} (t, {\bf {r}}) = \int _{0}^{+\infty} \sigma (t - t^{\prime}) {\bf E} (t^{\prime}, {\bf {r}}) dt^{\prime}\label{3.5} 
\end{equation}   
where   
\begin{equation}
\sigma (t - t^{\prime}) = \frac{\partial}{\partial t}\chi (t - t^{\prime})\label{3.6} 
\end{equation}
is a new memory function. A Fourier transformed $\sigma (t)$ defines the frequency-dependent complex conductivity of the medium, which is related to the susceptibility function via $\sigma (\omega) = -i\omega\chi(\omega)$. Causality requires that $\sigma (t - t^{\prime}) = 0$ for $t < t^{\prime}$ so that the integration in Eq.~(\ref{3.5}) is non-trivial only in the window $0\leq t^{\prime} \leq t$. Performing $\partial / \partial t$ on Eq.~(\ref{3.3}) we obtain the continuity equation
\begin{equation}
\frac{\partial}{\partial t} \rho (t, {\bf{r}}) + \nabla\cdot \textbf{j} (t, {\bf{r}}) = 0\label{3.7}
\end{equation}
with $\textbf{j} (t, {\bf{r}})$ in Eq.~(\ref{3.5}). In the following we proceed with the derivation of the KWW stretched exponential function.  

\subsection{KWW decay function and fractional relaxation equation}
By Laplace transforming Eqs.~(\ref{3.2}),~(\ref{3.5}), and~(\ref{3.7}) we get  
\begin{equation}
\nabla\cdot {\bf E} (\mathrm{s}, {\bf {r}}) = 4\pi\rho (\mathrm{s}, {\bf {r}})\label{3.2L}
\end{equation}
\begin{equation}
{\bf j} (\mathrm{s}, {\bf {r}}) = \sigma (\mathrm{s}) {\bf E} (\mathrm{s}, {\bf {r}})\label{3.5L}
\end{equation}
\begin{equation}
\mathrm{s}\rho (\mathrm{s}, {\bf{r}}) - \rho (0, {\bf{r}})  + \nabla\cdot \textbf{j} (\mathrm{s}, {\bf{r}}) = 0\label{3.7L} 
\end{equation}
If we apply $\nabla\cdot$ on Eq.~(\ref{3.5L}) and utilize Eq.~(\ref{3.2L}) we find 
\begin{equation}
\nabla\cdot\textbf{j} (\mathrm{s}, {\bf{r}}) = \sigma (\mathrm{s}) \nabla\cdot {\textbf E} (\mathrm{s}, {\bf{r}}) = 4\pi\sigma (\mathrm{s})\rho (\mathrm{s}, {\bf{r}})\label{3.8} 
\end{equation}
When substituted into Eq.~(\ref{3.7L}) this yields
\begin{equation}
\mathrm{s}\rho (\mathrm{s}, {\bf{r}}) + 4\pi\sigma (\mathrm{s})\rho (\mathrm{s}, {\bf{r}}) = \rho (0, {\bf{r}})\label{3.9}
\end{equation}
We assume that the sample observes ac universality, i.e., $\sigma (\mathrm{s}) = \alpha \mathrm{s} ^\eta$ with $\alpha$ a constant and $\eta$ a fraction between 0 and 1. This power-law form is just the Laplace version of Eq.~(\ref{1.3}). We shall use this form as an approximation of $\sigma (\mathrm{s})$ for the corresponding (short) time scales, $t \sim 1/ \mathrm{s}$. Equation~(\ref{3.9}) becomes   
\begin{equation}
\mathrm{s}\rho (\mathrm{s}, {\bf{r}}) + 4\pi \alpha \mathrm{s} ^\eta \rho (\mathrm{s}, {\bf{r}}) = \rho (0, {\bf{r}})\label{3.10}
\end{equation}
Separating variables we write $\rho (\mathrm{s}, {\bf{r}}) = \phi (\mathrm{s})\psi({\bf{r}})$ with $\phi (\mathrm{s})$ the Laplace transform of the relaxation function $\phi (t)$. From Eq.~(\ref{3.10}) it follows that
\begin{equation}
\phi (\mathrm{s}) = \frac{1}{\mathrm{s} + 4\pi \alpha \mathrm{s} ^\eta}\label{3.11}
\end{equation}
where the initial condition $\phi (0) = 1$ has been applied. In the time domain, 
\begin{equation}
\phi (t) = \frac{1}{2\pi i} \int _{-i\infty}^{+i\infty} \frac{e^{\mathrm{s} t}}{\mathrm{s} + \tau ^{-\beta}\,\mathrm{s} ^{1-\beta}} d\mathrm{s}\label{3.12}
\end{equation}
where we used the notations $\beta = 1-\eta$ and $\tau ^{-\beta} = 4\pi \alpha$. Equation~(\ref{3.12}) is exactly the definition of the Mittag-Leffler function $\mathrm{E} _\beta \left[-(t/\tau) ^{\beta}\right]$ (Eq. (B.1) in Appendix B of Ref. \onlinecite{Klafter}). The Mittag-Leffler function has the series expansion
\begin{equation}
\mathrm{E} _\beta \left[-(t/\tau) ^{\beta}\right] = \sum _{n=0}^{\infty} (-1)^n \frac{(t/\tau) ^{n\beta}}{\Gamma (n\beta + 1)}\label{3.13}
\end{equation}
For short times, this expansion goes as a stretched exponential, i.e.,
\begin{equation}
\mathrm{E} _\beta \left[-(t/\tau) ^{\beta}\right] \approx \exp \left[-\frac{(t/\tau)^\beta}{\Gamma (\beta + 1)}\right]\label{3.14}
\end{equation} 
This closed analytic form replicates the KWW relaxation function in Eq. (1). 

We now derive a dynamical relaxation equation from the dispersion relation~(\ref{3.10}). The key step is to notice that the power-law $\mathrm{s} ^\eta$ with $0<\eta < 1$ is the Laplace transformed Riemann-Liouville derivative, \cite{Klafter} which is defined through  
\begin{equation}
_{0} \mathrm{D}_t^{\eta} f (t, {\bf{r}}) = \frac{1}{\Gamma (1-\eta)} \frac{\partial}{\partial t} \int _{0}^{t} dt^{\prime} \frac{f (t^{\prime}, {\bf{r}})}{(t-t^{\prime})^{\eta}}\label{Riemann}
\end{equation}
The Riemann-Liouville derivative differs from the Riesz derivative in that the integration starts from $t=0$ and not from $t= -\infty$. Despite some particularities \cite{Klafter} of the composition rules and initial-value terms both vehicles share the property of being well-defined fractional extensions of the ordinary differentiation. In the limit of $\eta \rightarrow 1$, the fractional derivative in Eq.~(\ref{Riemann}) reduces to ordinary time derivative, $\partial / \partial t$. Replacing $\mathrm{s} ^\eta$ by $_{0} \mathrm{D}_t^{\eta}$ in Eq.~(\ref{3.10}) we write, with $\tau ^{-\beta} = 4\pi \alpha$, 
\begin{equation}
\frac{\partial}{\partial t} \rho (t, {\bf{r}}) = -\tau ^{-\beta}\, {_0} \mathrm{D}_t^{1 - \beta} \rho (t, {\bf{r}})\label{Frelax}
\end{equation}
It is instructive to derive Eq.~(\ref{Frelax}) directly from Eq.~(\ref{3.1}). Applying $\nabla\cdot$ to Eq.~(\ref{3.1}) and utilizing Eqs.~(\ref{3.2}) and~(\ref{3.3}) we find, with $\rho (0, {\bf {r}}) = -\nabla\cdot {\bf P} (0, {\bf {r}})$,
\begin{equation}
\rho (t, {\bf {r}}) = \rho (0, {\bf {r}}) - 4\pi \int _{0}^{t} \chi (t - t^{\prime}) \rho (t^{\prime}, {\bf {r}}) dt^{\prime}\label{3.15a} 
\end{equation}
which is a closed integral equation for $\rho (t, {\bf {r}})$. Note that the causality condition $\chi (t - t^{\prime}) = 0$ for $t < t^{\prime}$ has allowed us to set the upper limit of integration to $t$. By time differentiating Eq.~(\ref{3.15a}) we get
\begin{equation}
\frac{\partial}{\partial t}{\rho} (t, {\bf {r}}) =  - 4\pi \frac{\partial}{\partial t} \int _{0}^{t} \chi (t - t^{\prime}) \rho (t^{\prime}, {\bf {r}}) dt^{\prime}\label{3.15} 
\end{equation}
The memory kernel $\chi (t)$ is calculated from Eqs.~(\ref{2.3}) and~(\ref{3.12}). Here we are interested in the short-time behavior, which can be most readily evaluated from the power expansion in Eq.~(\ref{3.13}) to give
\begin{equation}
\chi (t)\propto \frac{\alpha}{\Gamma (\beta)}\,t^{\beta - 1}
\label{3.16}
\end{equation}
Of course, the scaling in Eq.~(\ref{3.16}) may be obtained as the inverse Laplace transform of the complex susceptibility, $\chi (\mathrm{s}) \propto \sigma (\mathrm{s}) / \mathrm{s} \propto \mathrm{s} ^{-\beta}$. When the power-law in Eq.~(\ref{3.16}) is substituted to Eq.~(\ref{3.15}), the Riemann-Liuville derivative ${_0} \mathrm{D}_t^{1 - \beta}$ is built, whence the dynamical Eq.~(\ref{Frelax}) follows.  

Equation~(\ref{Frelax}) with the Riemann-Liouville fractional derivative is the canonical form of the fractional relaxation equation. \cite{Klafter} The short-time behavior of the solution of the fractional relaxation equation is the KWW stretched-exponential decay function, in accordance with the power expansion of the Mittag-Leffler function.

Our main finding in Sec. III B is: The KWW stretched exponential decay function can be derived from the basic electrostatic equations under the additional assumption that ac conduction coefficient behaves as a fractional power of frequency.  

\subsection{Fractional diffusion equation describing sub-diffusion}
Our purpose now is to contrast the electrostatic description of the decay of polarization with a statistical-mechanical description of the dynamics of charged particles on the microscopic scales of the molecular motions. 

Microscopically, the relaxations are due to the motion of charges which interact with the fluctuating molecular environment. As a model approximation, we shall rely on the hypothesis of trap-controlled conduction and diffusion, in which the transport occurs as a result of hopping of charged-particles between the localized states. If the hopping has a characteristic time, then the transport is described by a Markovian chain process. In a more general situation there is a distribution of waiting or residence times between the consecutive steps of the motion and the Markovian property is invalidated. The system response to a charge-density perturbation is then a flow with memory: 
\begin{equation}
\textbf{j} (t, {\bf{r}}) = - \int _{0}^{t} \mathcal{D} (t - t^{\prime}) \nabla\rho (t^{\prime}, {\bf{r}}) dt^{\prime}\label{Fick}
\end{equation}
which goes against the concentration gradient as due to Fick's law. Here, $\mathcal{D} (t - t^{\prime})$ is a memory function, such that $\mathcal{D} (t - t^{\prime}) = 0$ for $t < t^{\prime}$. A Fourier transformed $\mathcal{D} (t)$ is defined as the frequency-dependent complex diffusion coefficient, $\mathcal{D} (\omega)$. The value of $\mathcal{D} (\omega)$ can be expressed in terms of the ac conduction coefficient as
\begin{equation}
\mathcal{D} (\omega) = \frac{T}{n e^2 }\sigma (\omega)\label{Einstein} 
\end{equation} 
where $e$ denotes the carrier charge, $n$ their number density, and $T$ the absolute temperature. In the zero-frequency limit, Eq.~(\ref{Einstein}) reduces to the conventional Einstein relation between the diffusion constant and the dc conductivity. Based on Eq.~(\ref{1.3}) we can argue that, $\mathcal{D} (\omega) \propto \omega ^{\eta}$ for the corresponding high frequencies. 

We now turn to demonstrate that, if the sample observes ac universality, the microscopic dynamics of charges is described by a fractional extension of the diffusion equation. 
 
By Laplace transforming Eq.~(\ref{Fick}) we get
\begin{equation}
\textbf{j} (\mathrm{s}, {\bf{r}}) = - \mathcal{D} (\mathrm{s}) \nabla\rho (\mathrm{s}, {\bf{r}})\label{3.17} 
\end{equation}
where $\mathcal{D} (\mathrm{s})$ is the Laplace transform of $\mathcal{D} (t)$. When substituted into the continuity Eq.~(\ref{3.7L}) this yields  
\begin{equation}
\mathrm{s}\rho (\mathrm{s}, {\bf{r}}) - \rho (0, {\bf{r}})  = \mathcal{D} (\mathrm{s}) \nabla ^2 \rho (\mathrm{s}, {\bf{r}})\label{3.18} 
\end{equation}
Adhering to the power-law form $\sigma (\mathrm{s}) = \alpha \mathrm{s} ^\eta$ from Eq.~(\ref{Einstein}) we have $\mathcal{D} (\mathrm{s}) = \Lambda \mathrm{s} ^{\eta}$ with $\Lambda = \alpha T/ne^2$. Utilizing this in Eq.~(\ref{3.18}) we write
\begin{equation}
\mathrm{s}\rho (\mathrm{s}, {\bf{r}}) - \rho (0, {\bf{r}})  = \mathrm{s} ^{\eta} \nabla ^2 \rho (\mathrm{s}, {\bf{r}})\label{3.18a} 
\end{equation}
where we set $\Lambda = 1$ for simplicity. In the time domain, Eq.~(\ref{3.18a}) reads, with $\eta = 1-\beta$,
\begin{equation}
\frac{\partial}{\partial t} \rho (t, {\bf{r}}) = {_0} \mathrm{D}_t^{1-\beta} \nabla^2 \rho (t, {\bf{r}})\label{FD} 
\end{equation}
Equation~(\ref{FD}) is the canonical form of the fractional diffusion equation describing sub-diffusion, \cite{Klafter} with $\beta$ the fractal dimension in time. \cite{PhysicaD} In various settings, this equation has been derived and discussed in the literature. \cite{Klafter,PhysicaD,Zaslavsky,Rest,UFN,Coffey} 

The characteristic function of the fractional diffusion Eq.~(\ref{FD}) obeys the fractional relaxation equation
\begin{equation}
\frac{\partial}{\partial t} \rho (t, {\bf{k}}) = -{\bf k} ^2 \, {_0} \mathrm{D}_t^{1-\beta} \rho (t, {\bf{k}})\label{Char} 
\end{equation}
where ${\bf{k}}$ is the wave-vector in the ambient real-space. From Eq.~(\ref{3.14}) one can see that the short-time behavior of $\rho (t, {\bf{k}})$ is the stretched exponential  
\begin{equation}
\rho (t, {\bf{k}}) \approx \exp \left[-\frac{{\bf{k}}^2 t ^{\beta}}{\Gamma (\beta + 1)}\right]
\end{equation}
With this observation we conclude the analysis of connection between the KWW stretched exponential relaxation function and ac universality in disordered solids.

\section{Summary}
We have analyzed the properties of dielectric relaxation and ac (alternating-current) conduction in disordered dielectrics. We have discussed the route to the statistical mechanical foundation and suggested physical models that might connect to the typically observed dynamical characteristics. We proposed a self-consistent model of dielectric relaxation, in which the polarization and electric source fields are self-consistently generated by the residual polarization-charge density. This self-consistent approach has led us to a systematic derivation of the fractional relaxation and fractional diffusion equations from the property of ac universality. Our results support the hypothesis that dielectrics exhibiting ac universality and stretched exponential relaxations are in a self-organized critical state.   


\begin{acknowledgments} 
This work was supported in full under the project No 171076/V30 of the Norwegian Research Council.
\end{acknowledgments}


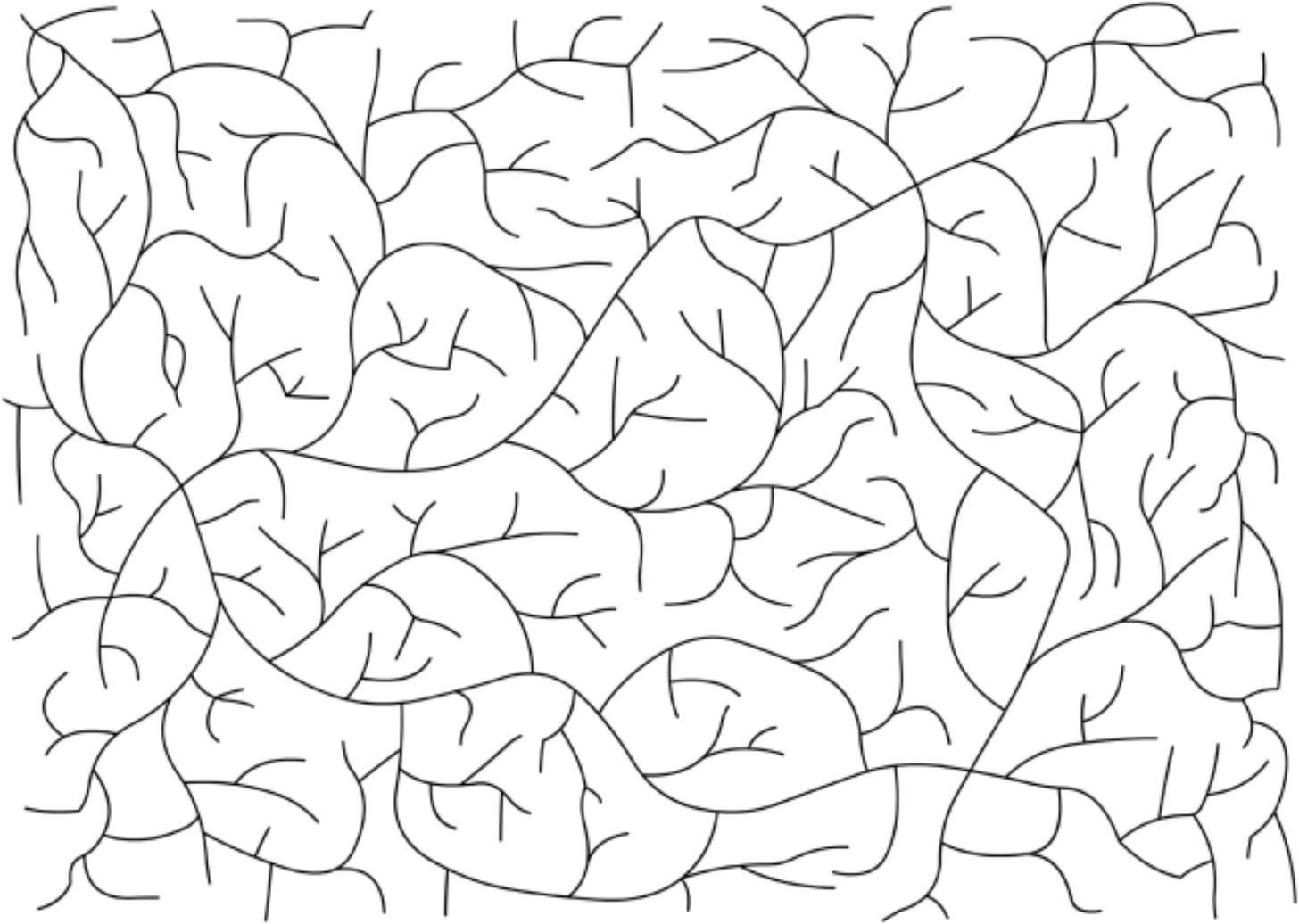
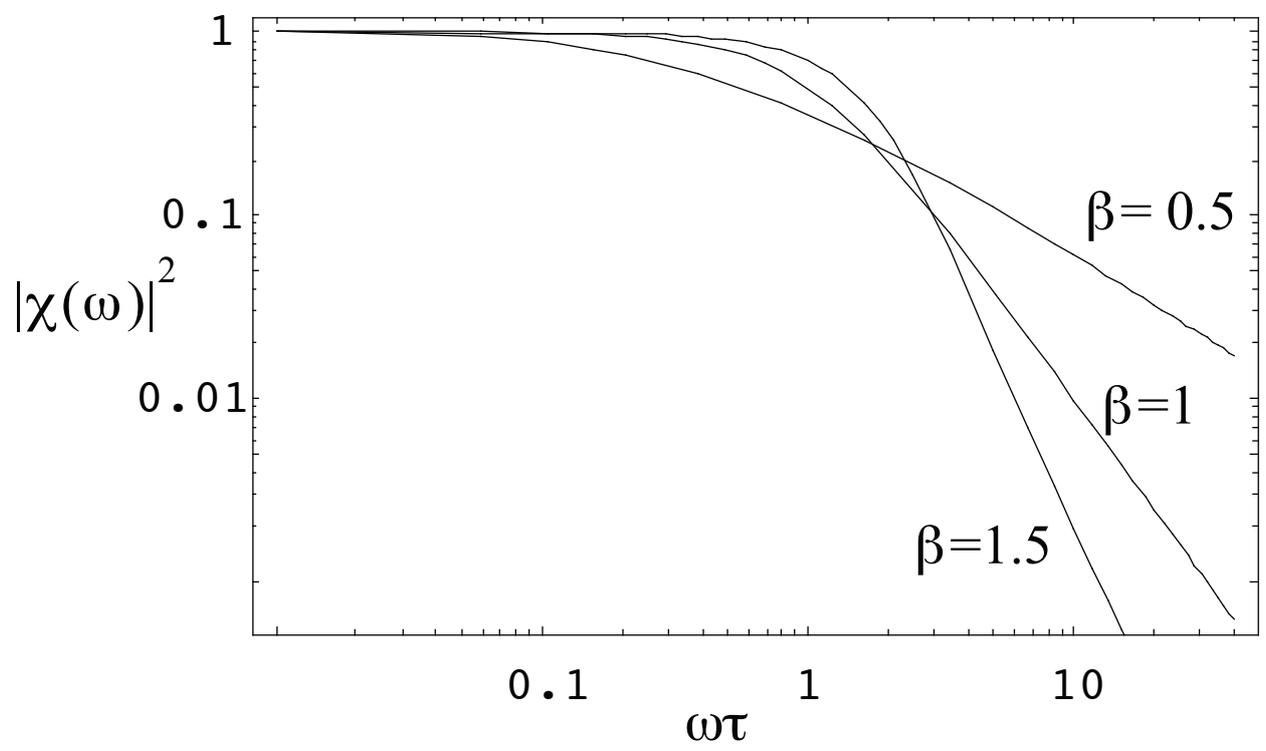
\end{document}